\documentclass[12pt]{amsproc}%
\usepackage{amssymb}
\usepackage{amsmath}
\usepackage{amsfonts}
\usepackage{graphicx}%
\begin{document}
\title{Power-law Behavior of High Energy String Scatterings in Compact Spaces}
\author{Jen-Chi Lee}
\address{Department of Electrophysics, National Chiao-Tung University, Hsinchu, Taiwan, R.O.C.}
\email{jcclee@cc.nctu.edu.tw }
\author{Yi Yang}
\address{Department of Electrophysics, National Chiao-Tung University and Physics
Division, National Center for Theoretical Sciences, Hsinchu, Taiwan, R.O.C.}
\email{yyang@phys.cts.nthu.edu.tw}
\date{\today }

\begin{abstract}
We calculate high energy massive scattering amplitudes of closed bosonic
string compactified on the torus. We obtain infinite linear relations among
high energy scattering amplitudes. For some kinematic regimes, we discover
that some linear relations break down and, simultaneously, the amplitudes
enhance to power-law behavior due to the space-time T-duality symmetry in the
compact direction. This result is consistent with the coexistence of the
linear relations and the softer exponential fall-off behavior of high energy
string scattering amplitudes as we pointed out prevously. It is also
reminiscent of hard (power-law) string scatterings in warped spacetime
proposed by Polchinski and Strassler.

\end{abstract}
\maketitle

\section{\bigskip Introduction and Overview}

It is well known that there are two fundamental characteristics of high energy
string scattering amplitudes, which make them very different from field theory
scatterings. These are the softer exponential fall-off behavior (in contrast
to the hard power-law behavior of field theory scatterings) and the existence
of infinite Regge-pole structure in the form factor of the high energy string
scattering amplitudes.

For the last few years, high-energy, fixed angle behavior of string scattering
amplitudes \cite{GM, Gross, GrossManes} was intensively reinvestigated for
massive string states at arbitrary mass levels \cite{ChanLee1,ChanLee2,
CHL,CHLTY,PRL,paperB,susy,Closed,HL}. An infinite number of linear relations
among string scattering amplitudes of different string states were discovered.
An important new ingredient of these calculations is the zero-norm states
(ZNS) \cite{ZNS1,ZNS3,ZNS2} in the old covariant first quantized (OCFQ) string
spectrum. The discovery of these infinite linear relations constitutes the
\textit{third} fundamental characteristics of high energy string scatterings,
which is not shared by the usual point-particle field theory scatterings.

More recently, it was tempted to conjecture that \cite{Dscatt, Wall, Decay}
the newly discovered linear relations, or stringy symmetries, are responsible
for the softer exponential fall-off string scatterings at high energies. One
way to justify this conjecture (that is: the coexistence of the infinite
linear relations and the softer exponential fall-off behavior of high energy
string scatterings) is to find more examples of high energy string
scatterings, which show the unusual hard power-law behavior and,
simultaneously, give the breakdown of the infinite linear relations. With this
in mind, in this report \cite{Compact} we calculate high energy, fixed angle
massive scattering amplitudes of closed bosonic string compactified on the
torus \cite{Mende}. In the Gross regime (GR), for each fixed mass level with
given quantized and winding momenta $\left(  \frac{m}{R},\frac{1}{2}nR\right)
$, we obtain infinite linear relations among high energy scattering amplitudes
of different string states. Moreover we discover that, for some kinematic
regime, the so called Mende regime (MR), infinite linear relations with
$N_{R}=N_{L}$ break down and, simultaneously, the amplitudes enhance to
power-law behavior \cite{Compact}. It is the space-time T-duality symmetry
that plays a role here.

There was another motivation to study the unusual high energy hard power-law
behavior of string scattering. This is mainly motivated by the Gauge/String
duality in the Type II B string theory on $AdS_{5}$ background
\cite{Maldacena}. The work of Polchinski and Strassler and others \cite{PS,
And} suggested that the high energy behavior of string scattering in warped
spacetime gives a consistent hard power-law behavior. It would be an
interesting problem to understand the common features of the power-law string
scatterings in these two different string backgrounds.

\section{\bigskip High Energy Scattering}

We consider 26D closed bosonic string with one coordinate compactified on
$S^{1}$ with radius $R$. The closed string boundary condition for the
compactified coordinate is%

\begin{equation}
X^{25}(\sigma+2\pi,\tau)=X^{25}(\sigma,\tau)+2\pi Rn,
\end{equation}
where $n$ is the winding number. The momentum in the $X^{25}$ direction is
then quantized to be%
\begin{equation}
K=\frac{m}{R},
\end{equation}
where $m$ is an integer. The left and right momenta are defined to be%
\begin{equation}
K_{L,R}=K\pm L=\frac{m}{R}\pm\dfrac{1}{2}nR\Rightarrow K=\dfrac{1}{2}\left(
K_{L}+K_{R}\right)  ,
\end{equation}
and the mass spectrum can be calculated to be%
\begin{equation}
\left\{
\begin{array}
[c]{c}%
M^{2}=\left(  \dfrac{m^{2}}{R^{2}}+\dfrac{1}{4}n^{2}R^{2}\right)  +N_{R}%
+N_{L}-2\equiv K_{L}^{2}+M_{L}^{2}\equiv K_{R}^{2}+M_{R}^{2}\\
N_{R}-N_{L}=mn
\end{array}
\right.  , \label{mass}%
\end{equation}
where $N_{R}$ and $N_{L}$ are the number operators for the right and left
movers, which include the counting of the compactified coordinate. We have
also introduced the left and the right level masses as%
\begin{equation}
M_{L,R}^{2}\equiv2\left(  N_{L,R}-1\right)  . \label{level mass}%
\end{equation}

In the center of momentum frame, the kinematic can be set up to be%

\begin{align}
k_{1L,R}  &  =\left(  +\sqrt{p^{2}+M_{1}^{2}},-p,0,-K_{1L,R}\right)  ,\\
k_{2L,R}  &  =\left(  +\sqrt{p^{2}+M_{2}^{2}},+p,0,+K_{2L,R}\right)  ,\\
k_{3L,R}  &  =\left(  -\sqrt{q^{2}+M_{3}^{2}},-q\cos\phi,-q\sin\phi
,-K_{3L,R}\right)  ,\\
k_{4L,R}  &  =\left(  -\sqrt{q^{2}+M_{4}^{2}},+q\cos\phi,+q\sin\phi
,+K_{4L,R}\right)
\end{align}
where $p\equiv\left\vert \mathrm{\vec{p}}\right\vert $ and $q\equiv\left\vert
\mathrm{\vec{q}}\right\vert $ and%
\begin{align}
k_{i}  &  \equiv\dfrac{1}{2}\left(  k_{iR}+k_{iL}\right)  ,\\
k_{i}^{2}  &  =K_{i}^{2}-M_{i}^{2},\\
k_{iL,R}^{2}  &  =K_{iL,R}^{2}-M_{i}^{2}\equiv-M_{iL,R}^{2}.
\end{align}
With this setup, the center of mass energy $E$ is%
\begin{equation}
E=\dfrac{1}{2}\left(  \sqrt{p^{2}+M_{1}^{2}}+\sqrt{p^{2}+M_{2}^{2}}\right)
=\dfrac{1}{2}\left(  \sqrt{q^{2}+M_{3}^{2}}+\sqrt{q^{2}+M_{4}^{2}}\right)  .
\label{COM}%
\end{equation}
The conservation of momentum on the compactified direction gives%
\begin{equation}
m_{1}-m_{2}+m_{3}-m_{4}=0, \label{kk}%
\end{equation}
and T-duality symmetry implies conservation of winding number%
\begin{equation}
n_{1}-n_{2}+n_{3}-n_{4}=0. \label{wind}%
\end{equation}
The left and the right Mandelstam variables are defined to be%
\begin{align}
s_{L,R}  &  \equiv-(k_{1L,R}+k_{2L,R})^{2},\\
t_{L,R}  &  \equiv-(k_{2L,R}+k_{3L,R})^{2},\\
u_{L,R}  &  \equiv-(k_{1L,R}+k_{3L,R})^{2}.
\end{align}

We now proceed to calculate the high energy scattering amplitudes for general
higher mass levels with fixed $N_{R}+N_{L}$. With one compactified coordinate,
the mass spectrum of the second vertex of the amplitude is%
\begin{equation}
M_{2}^{2}=\left(  \dfrac{m_{2}^{2}}{R^{2}}+\dfrac{1}{4}n_{2}^{2}R^{2}\right)
+N_{R}+N_{L}-2.
\end{equation}
We now have more mass parameters to define the "high energy limit". We are
going to use three quantities $E^{2},M_{2}^{2}$ and $N_{R}+N_{L}$ to define
different regimes of "high energy limit". The high energy regime defined by
$E^{2}\simeq M_{2}^{2}$ $\gg$ $N_{R}+N_{L}$ will be called Mende regime (MR).
The high energy regime defined by $E^{2}\gg M_{2}^{2}$, $E^{2}\gg$
$N_{R}+N_{L}$ will be called Gross region (GR). In the high energy limit, the
polarizations on the scattering plane for the second vertex operator are
defined to be%

\begin{align}
e^{\mathbf{P}} &  =\frac{1}{M_{2}}\left(  \sqrt{p^{2}+M_{2}^{2}},p,0,0\right)
,\\
e^{\mathbf{L}} &  =\frac{1}{M_{2}}\left(  p,\sqrt{p^{2}+M_{2}^{2}},0,0\right)
,\\
e^{\mathbf{T}} &  =\left(  0,0,1,0\right)
\end{align}
where the fourth component refers to the compactified direction. In the MR, we
will use \cite{Compact}%
\begin{equation}
\left\vert N_{L,R},q_{L,R}\right\rangle \equiv\left(  \alpha_{-1}^{\mathbf{T}%
}\right)  ^{N_{L}-2q_{L}}\left(  \alpha_{-2}^{\mathbf{P}}\right)  ^{q_{L}%
}\otimes\left(  \tilde{\alpha}_{-1}^{\mathbf{T}}\right)  ^{N_{R}-2q_{R}%
}\left(  \tilde{\alpha}_{-2}^{\mathbf{P}}\right)  ^{q_{R}}\left\vert
0\right\rangle \label{new}%
\end{equation}
as the second vertex operator in the calculation of high energy scattering
amplitudes. The high energy scattering amplitudes in the MR can be calculated
to be%
\begin{align}
A &  \simeq\left(  -\dfrac{q\sin\phi\left(  s_{L}+t_{L}\right)  }{t_{L}%
}\right)  ^{N_{L}}\left(  -\dfrac{q\sin\phi\left(  s_{R}+t_{R}\right)  }%
{t_{R}}\right)  ^{N_{R}}\left(  \frac{1}{2M_{2}q^{2}\sin^{2}\phi}\right)
^{q_{L}+q_{R}}\nonumber\\
&  \cdot\left(  \left(  t_{R}-2\vec{K}_{2R}\cdot\vec{K}_{3R}\right)
+\dfrac{t_{R}^{2}\left(  s_{R}-2\vec{K}_{1R}\cdot\vec{K}_{2R}\right)  }%
{s_{R}^{2}}\right)  ^{q_{R}}\nonumber\\
&  \cdot\left(  \left(  t_{L}-2\vec{K}_{2L}\cdot\vec{K}_{3L}\right)
+\dfrac{t_{L}^{2}\left(  s_{L}-2\vec{K}_{1L}\cdot\vec{K}_{2L}\right)  }%
{s_{L}^{2}}\right)  ^{q_{L}}\nonumber\\
&  \cdot\frac{\sin\left(  \pi s_{L}/2\right)  \sin\left(  \pi t_{R}/2\right)
}{\sin\left(  \pi u_{L}/2\right)  }B\left(  -1-\dfrac{t_{R}}{2},-1-\dfrac
{u_{R}}{2}\right)  B\left(  -1-\dfrac{t_{L}}{2},-1-\dfrac{u_{L}}{2}\right)
.\label{amplitude}%
\end{align}
Eq.(\ref{amplitude}) is valid for $E^{2}\gg N_{R}+N_{L},$ $M_{2}^{2}\gg
N_{R}+N_{L}.$

\subsection{The infinite linear relations in the GR}

\bigskip For the special case of GR with $E^{2}\gg M_{2}^{2}$,
Eq.(\ref{amplitude}) can be further reduced to%
\begin{align}
\lim_{E^{2}\gg M_{2}^{2}}A &  \simeq\left(  -\frac{2\cot\frac{\phi}{2}}%
{E}\right)  ^{N_{L}+N_{R}}\left(  -\frac{1}{2M_{2}}\right)  ^{q_{L}+q_{R}%
}E^{-1}\left(  \sin\frac{\phi}{2}\right)  ^{-3}\left(  \cos\frac{\phi}%
{2}\right)  ^{5}\nonumber\\
&  \cdot\frac{\sin\left(  \pi s_{L}/2\right)  \sin\left(  \pi t_{R}/2\right)
}{\sin\left(  \pi u_{L}/2\right)  }\exp\left(  -\frac{t\ln t+u\ln
u-(t+u)\ln(t+u)}{4}\right)  .\label{linear}%
\end{align}
We see that, in the GR, for each fixed mass level with given quantized and
winding momenta $\left(  \frac{m}{R},\frac{1}{2}nR\right)  $, we have obtained
infinite linear relations among high energy scattering amplitudes of different
string states with various $(q_{L},q_{R})$. Note also that this result
reproduces the correct ratios $\left(  -\frac{1}{2M_{2}}\right)  ^{q_{L}%
+q_{R}}$ obtained in the previous works \cite{Dscatt, Wall, Decay}. However,
the mass parameter $M_{2}$ here depends on $\left(  \frac{m}{R},\frac{1}%
{2}nR\right)  $.

\subsection{Power-law and breakdown of the infinite linear relations in the
MR}

\bigskip The power-law behavior of high energy string scatterings in a compact
space was first suggested by Mende. Here we give a mathematically more
concrete description. It is easy to see that the "power law" condition, i.e.
Eq.(3.7) in Mende's paper \cite{Mende}%
\begin{equation}
k_{1L}\cdot k_{2L}+k_{1R}\cdot k_{2R}=\text{constant,}%
\label{mandy's condition}%
\end{equation}
turns out to be%
\begin{align}
&  \sqrt{p^{2}+M_{1}^{2}}\cdot\sqrt{p^{2}+M_{2}^{2}}+p^{2}+2\left(  \vec
{K}_{1}\cdot\vec{K}_{2}+\vec{L}_{1}\cdot\vec{L}_{2}\right)  \nonumber\\
&  =\text{constant.}%
\end{align}
As $p\rightarrow\infty$, due to the existence of winding modes in the
compactified closed string, it is possible to choose $\left(  \vec{K}_{1}%
,\vec{K}_{2};\vec{L}_{1},\vec{L}_{2}\right)  $\ such that%
\begin{equation}
\vec{K}_{1}\cdot\vec{K}_{2}+\vec{L}_{1}\cdot\vec{L}_{2}<0,
\end{equation}
and let $\left(  \vec{K}_{1}\cdot\vec{K}_{2}+\vec{L}_{1}\cdot\vec{L}%
_{2}\right)  \rightarrow-\infty$ to make
\begin{align}
k_{1L}\cdot k_{2L}+k_{1R}\cdot k_{2R}\simeq &  \text{ constant}\\
\Rightarrow s_{L}+s_{R}\simeq &  \text{ constant.}%
\end{align}
In our calculation, this condition implies the beta functions in
Eq.(\ref{amplitude}) reduce to%
\begin{align}
&  B\left(  -1-\dfrac{t_{R}}{2},-1-\dfrac{u_{R}}{2}\right)  B\left(
-1-\dfrac{t_{L}}{2},-1-\dfrac{u_{L}}{2}\right)  \nonumber\\
&  =\frac{\sin\left(  \pi s_{R}/2\right)  \Gamma(-\frac{t_{R}}{2}%
-1)\Gamma(-\frac{u_{R}}{2}-1)\Gamma(-\frac{t_{L}}{2}-1)\Gamma(-\frac{u_{L}}%
{2}-1)}{\pi\frac{s_{R}}{2}\left(  1+\frac{s_{R}}{2}\right)  \left(
-1+\frac{s_{R}}{2}\right)  },
\end{align}
which behaves as \textit{power-law} in the high energy limit! On the other
hand, it is obvious that the $(q_{L},q_{R})$ dependent power factors of the
amplitude in Eq.(\ref{amplitude})%
\begin{align}
A_{q_{L},q_{R}} &  \simeq\left(  \frac{1}{2M_{2}q^{2}\sin^{2}\phi}\right)
^{q_{L}+q_{R}}\nonumber\\
&  \cdot\left(  \left(  t_{R}-2\vec{K}_{2R}\cdot\vec{K}_{3R}\right)
+\dfrac{t_{R}^{2}\left(  s_{R}-2\vec{K}_{1R}\cdot\vec{K}_{2R}\right)  }%
{s_{R}^{2}}\right)  ^{q_{R}}\nonumber\\
&  \cdot\left(  \left(  t_{L}-2\vec{K}_{2L}\cdot\vec{K}_{3L}\right)
+\dfrac{t_{L}^{2}\left(  s_{L}-2\vec{K}_{1L}\cdot\vec{K}_{2L}\right)  }%
{s_{L}^{2}}\right)  ^{q_{L}}%
\end{align}
show \textit{no} linear relations in the MR. Note that the mechanism to break
the linear relations and the mechanism to enhance the amplitude to power-law
are all due to $E\simeq M_{2}$ in the MR. In our notation,
Eq.(\ref{mandy's condition}) is equivalent to the following condition%
\begin{equation}
\lim_{p\rightarrow\infty}\frac{\sqrt{p^{2}+M_{1}^{2}}\cdot\sqrt{p^{2}%
+M_{2}^{2}}+p^{2}}{\vec{K}_{1}\cdot\vec{K}_{2}+\vec{L}_{1}\cdot\vec{L}_{2}%
}\sim\frac{E^{2}}{\left(  \dfrac{m_{1}m_{2}}{R^{2}}+\dfrac{1}{4}n_{1}%
n_{2}R^{2}\right)  }\sim-\text{ }\mathcal{O}(1).\label{condition}%
\end{equation}
For our purpose here, as we will see soon, it is good enough to choose only
one compactified coordinate to realize Eq.(\ref{condition}). First of all, in
addition to Eq.(\ref{kk}) and Eq.(\ref{wind}), Eq.(\ref{mass}) implies%
\begin{equation}
m_{i}n_{i}=0,i=1,2,3,4\text{ (no sum on }i\text{).}%
\end{equation}
This is because three of the four vertex are tachyons. Also, since we are
going to take $n_{2}$ to infinity with fixed $N_{R}+N_{L}$ in order to satisfy
Eq.(\ref{condition}), we are forced to take $m_{2}=0$. In sum, we can take,
say, $m_{i}=0$ for $i=1,2,3,4,$ and $n_{1}=-n_{2}=-n,n_{3}=-2n,n_{4}=0,$ and
then let $n\rightarrow\infty$ to realize Eq.(\ref{condition}). Note that it is
crucial to choose different sign for $n_{1}$ and $n_{2}$ in order to achieve
the minus sign in Eq.(\ref{condition}). We stress that there are other choices
to realize the condition. One notes that all choices implies%
\begin{equation}
N_{R}=N_{L}.
\end{equation}
It is obvious that one can also compactify more than one coordinate to realize
the Mende condition. We conclude that the high energy scatterings of the
"highly winding string states" of the compactified closed string in the MR
behave as the unusual UV power-law, and the usual linear relations among
scattering amplitudes break down due to the unusual power-law behavior.

This work is supported in part by the National Science Council, 50 billions
project of Ministry of Educaton and National Center for Theoretical Science,
Taiwan, R.O.C.

\end{document}